\documentstyle[12pt,epsfig]{article} 
\newcommand{\tr}{{\textrm {tr}}}

\newcommand{\D}{{\widehat D}}

\newcommand{\cL}{{\cal L}}

\def\slashchar#1{\setbox0=\hbox{$#1$}
   \dimen0=\wd0 \setbox1=\hbox{/} \dimen1=\wd1
   \ifdim\dimen0>\dimen1 \rlap{\hbox to \dimen0{\hfil/\hfil}} #1
   \else  \rlap{\hbox to \dimen1{\hfil$#1$\hfil}} / \fi}

\def\D{{\bf D}}

\begin{document}

\title{ Polyakov Loop at Finite Temperature in Chiral Quark
Models\footnote{ Talk given by E.R.A. at the Miniworkshop on ``Quark
Dynamics'', Bled, Slovenia, 7-12 July-2004.}}
\author{E. Meg\'{\i}as\footnote{email:emegias@ugr.es}, E. Ruiz
Arriola\footnote{email:earriola@ugr.es}, and
L.L. Salcedo\footnote{email:salcedo@ugr.es}\\ Departamento de
F\'{\i}sica Moderna, \\ Universidad de Granada, E-18071 Granada, Spain
} \date{\today} \maketitle
\begin{abstract}
At finite temperature, chiral quark models do not incorporate large
gauge invariance which implies genuinely non-perturbative finite
temperature gluonic degrees of freedom. Motivated by this observation,
we describe how the coupling of the Polyakov loop as an independent
degree of freedom to quarks not only accounts for large gauge
invariance, but also allows to establish in a dynamical way the
interaction between composite hadronic states such as Goldstone bosons
to finite temperature non-perturbative gluons in a medium which can
undergo a confinement-deconfinement phase transition.
\end{abstract}

\section{Large Gauge Transformations}

One feature of gauge theories like QCD at finite temperatures in the
imaginary time formulation~\cite{Gross:1980br,Landsman:1986uw,
Svetitsky:1985ye} is the non-perturbative manifestation of the non
Abelian gauge symmetry. In the Polyakov gauge, where $ \partial_4
A_4=0$ and $ A_4 $ is a diagonal and traceless $N_c \times N_c $
matrix, and $N_c$ is the number of colors, there is still some freedom
in choosing the gluon field. Let us consider for instance the periodic
gauge transformation \cite{Salcedo:1998sv,Garcia-Recio:2000gt}
\begin{eqnarray}
g (x_4 ) = e^{i 2 \pi x_4  \Lambda/\beta} \, , 
\label{eq:lgt}
\end{eqnarray}
where $\Lambda $ is a color traceless diagonal matrix of integers. We
call it a large gauge transformation (LGT) since it cannot be
considered to be close to the identity\footnote{Note that they are not
large in the topological sense, as discussed in
\cite{Salcedo:1998sv,Garcia-Recio:2000gt}.}. The gauge
transformation on the $A_4$ component of the gluon field is
\begin{eqnarray}
A_4 \to A_4 + \frac{2 \pi}{ \beta} \Lambda \,.
\end{eqnarray} 
Thus, invariance under the LGT, Eq.~(\ref{eq:lgt}), implies a constant
shift in the $A_4$ gluon amplitudes, meaning that $A_4$ is not
uniquely defined by the Polyakov gauge condition.  These ambiguities
on the choice of the gauge field within a given gauge fixing are
usually called Gribov copies. The requirement of gauge invariance
actually implies identifying all amplitudes differing by a multiple of
$2 \pi / \beta $, which means periodicity in the diagonal amplitudes
of $A_4$ of period $ 2 \pi / \beta$. Perturbation theory, which
corresponds to expanding in powers of small $A_4$ fields manifestly
breaks gauge invariance at finite temperature, because a Taylor
expansion on a periodic function violates the periodicity
behavior. Thus, taking into account these Gribov replicas is
equivalent to explicitly deal with genuine non-perturbative finite
temperature gluonic degrees of freedom. A way of automatically taking
into account LGT is by considering the Polyakov loop $\Omega$ as an
independent variable, which in the Polyakov gauge becomes a
diagonal unitary matrix
\begin{eqnarray}
\Omega  = e^{{\rm i} \beta A_4 (\vec x)}
\end{eqnarray} 
invariant under the set of transformations given by
Eq.~(\ref{eq:lgt}).  The relevance of the Polyakov loop in practical
calculations is well recognized~\cite{Gross:1980br} but seldomly taken
into account in high temperature calculations where large gauge
invariance is manifestly broken since the gluon field is considered to
be small. We have recently developed an expansion keeping these
symmetries in general theories and applied it to QCD at the one
quark+gluon loop level~\cite{Megias:2002vr,Megias:2003ui}.

\section{The Center Symmetry}

In pure gluodynamics, or in the quenched approximation (valid for
heavy quarks) at finite temperature there is actually a larger
symmetry since one can extend the periodic transformations to
aperiodic ones~\cite{Svetitsky:1985ye},
\begin{eqnarray}
g(x_4 + \beta ) =  z g (x_4 ) \,,  \qquad z^{N_c} = 1  
\end{eqnarray}  
so that $z$ is an element of the center $Z(N_c)$ of the group $SU(N_c)
$. This center symmetry is a symmetry of the action as well as the
gluon field boundary conditions. An example of such a transformation
in the Polyakov gauge is given by
\begin{eqnarray}
g (x_4 ) = e^{i 2 \pi x_4  \Lambda/N_c\beta} \,.
\label{eq:ALGT} 
\end{eqnarray}
On the $A_4$ component of the  gluon field produces
\begin{eqnarray}
A_4 \to A_4 + \frac{2 \pi}{ N_c \beta } \Lambda \,.
\end{eqnarray} 
Thus, in the quenched approximation the period is $N_c$ times smaller
than in full QCD. Under these transformations the gluonic action,
measure and boundary conditions are invariant. The Polyakov loop,
however, transforms as the fundamental representation of the $Z(N_c)$
group, i.e.  $ \Omega \to z \Omega $, yielding $ \langle \Omega
\rangle = z \langle \Omega \rangle $ and hence $\langle \Omega \rangle
=0 $ in the unbroken center symmetry phase. At high temperatures one
expects perturbation theory to hold, the gluon field amplitude becomes
small and hence $\langle \Omega \rangle \to 1 $, justifying the choice
of $\Omega $ as an order parameter for a confinement-deconfinement
phase transition. More generally, in the confining phase
\begin{eqnarray}
\langle \Omega^n \rangle =0 \qquad {\rm for} \qquad n \neq m N_c
\label{eq:Pave}
\end{eqnarray} 
with $m$ an arbitrary integer. The antiperiodic quark fields at the
end of the Euclidean imaginary interval transform as $ q( \vec x ,
\beta ) = - q(\vec x, 0) \to z q( \vec x, \beta  ) = -  q (\vec x,0)$, so
that the center symmetry is explicitly broken by the presence of
dynamical quarks. A direct consequence of such a property is that, in
the quenched approximation non-local condensates fulfill a selection
rule of the form,
\begin{eqnarray}
\langle \bar q ( n \beta ) q( 0) \rangle =0 \qquad {\rm for}
\qquad n \neq m N_c
\label{eq:sel} 
\end{eqnarray} 
since under the large aperiodic transformations given by
Eq.~(\ref{eq:ALGT}) we have $ \bar q ( n \beta ) q( 0) \to z^{-n} \bar
q ( n \beta ) q( 0) $. This selection rule has some impact on chiral
quark models.

\section{Chiral quark models at finite temperature}

To fully appreciate the role played by the center symmetry in chiral
quark models~(for a recent review on such models see
e.g. Ref.~\cite{RuizArriola:2002wr} and references therein) let us
evaluate the chiral condensate at finite temperature. At the one loop
level one has\footnote{We use an asterisk to denote finite temperature
observables.}
\begin{eqnarray}
\langle \bar q q \rangle^* &=& 4 M T  {\rm Tr}_c \sum_{\omega_n}
\int \frac{d^3 k}{(2\pi)^3} \frac1{\omega_n^2+ k^2 + M^2 }  
\label{eq:cond}
\end{eqnarray} 
where $ \omega_n = 2\pi T ( n +1/2 ) $ are the fermionic Matsubara
frequencies, $M$ is the constituent quark mass and $ {\rm Tr}_c$
stands for the color trace in the fundamental representation which in
this case trivially yields a $N_c$ factor. Possible finite cut-off
corrections, appearing in the chiral quark models such as the NJL
model at finite temperature have been neglected. This is a reasonable
approximation as long as the temperature is low enough $ T \ll \Lambda
\sim 1 {\rm GeV} $. The condensate can be rewritten as
\begin{eqnarray}
\langle \bar q q \rangle^* &=& \sum_{n} (-1)^n \langle \bar q ( n
\beta ) q( 0) \rangle 
\label{eq:cond_njl}
\end{eqnarray} 
in terms of nonlocal Euclidean condensates at zero temperature. After
Poisson resummation, at low temperatures we have
\begin{eqnarray}
\langle \bar q q \rangle^* &=& \langle \bar q q \rangle + 8 N_c
\sum_{n = 1}^\infty (-)^n \frac{T M^2 }{\pi^2} K_1 ( M n /T ) 
\,,
\nonumber \\ & \sim & \langle \bar q q \rangle - \sum_{n = 1}^\infty
(-)^n \frac{N_c} 2 \left(\frac{2 M n T }{\pi} \right)^{3/2} e^{-n M /
T} \,,
\end{eqnarray} 
where the asymptotic expansion of the modified Bessel function $K_1$
has been used. One can interpret the previous formula for the
condensate in terms of statistical Boltzmann factors, since at large
Euclidean coordinates the fermion propagator behaves as $ S( i \beta,
\vec x ) \sim e^{-M \beta } $, so that we have contributions from
multiquark states. This is a problem since it means that the heat bath
is made out of free constituent quarks without any color
clustering\footnote{One could think that this is a natural consequence
of the lack of confinement in chiral quark models such as
NJL. Contrary to naive expectations this is not necessarily the case;
Boltzmann factors occur in quark models with analytic confinement such
as the Spectral Quark Model~\cite{RuizArriola:2003bs}. There the
condensate is given by
\begin{eqnarray} 
\frac{\langle \bar q q \rangle_*}{\langle \bar q q \rangle} &=& \tanh
(M/2T) = 1 -2 e^{-M/T} + 2 e^{-2M/T} + \dots
\label{eq:cond_SQM}
\end{eqnarray} 
where $M=M_S/2 $, despite the absence of poles in the quark
propagator.}. Another problem comes from comparison with Chiral
Perturbation Theory at Finite Temperature~\cite{Gasser:1986vb}. In the
chiral limit, i.e., for $ m_\pi \ll 2 \pi T \ll 4 \pi f_\pi $ the
leading thermal corrections to the quark condensate are given by
\begin{eqnarray}
\langle \bar q q \rangle^* \Big|_{\rm ChPT} &= & \langle \bar q q
\rangle \left( 1- \frac{T^2 } {8 f_\pi^2} - \frac{T^4 } {384 f_\pi^4}
+ \dots \right) \,.
\label{eq:chpt} 
\end{eqnarray} 
This formula is derived under the assumption that there is no
temperature dependence of the low energy constants, i.e. $ L_i^*
\simeq L_i $ so that the whole effect is due to thermal pion
loops. Thus, the finite temperature correction is $N_c$-suppressed as
compared to the zero temperature value. This {\it is not} what one
sees in chiral quark model calculations; in the large $N_c$ limit {\it
there is} a finite temperature correction, which would mean that the
low energy constants which appear in the chiral Lagrangian would have
a genuine tree level temperature dependence, $L_i^* - L_i \simeq N_c
e^{-M/T} $.  To obtain the ChPT result of Eq.~(\ref{eq:chpt}) pion
loops have to be considered~\cite{Florkowski:1996wf} and dominate for
$ T \ll M $. The problem is that already without pion loops chiral
quark models predict a chiral phase transition at about $T_c \sim
170$~MeV, in remarkable but perhaps unjustified agreement with lattice
calculations.

\section{Coupling the Polyakov loop}

In the Polyakov gauge one can formally keep track of large gauge
invariance at finite temperature by coupling gluons to the model in a
minimal way. This means in practice using the modified fermionic
Matsubara frequencies \cite{Salcedo:1998sv,Garcia-Recio:2000gt}
\begin{eqnarray}
\hat \omega_n = 2 \pi T ( n+1/2 + \nu)\,, \quad \nu=(2\pi i)^{-1}\log\Omega
\end{eqnarray}
which are shifted by the logarithm of the Polyakov loop which we
assume for simplicity to be $\vec x$ independent. Previous work have
coupled similarly $\Omega$ on pure phenomenological
grounds~\cite{Gocksch:1984yk,Meisinger:2002kg,Fukushima:2003fw}, but
the key role played by the implementation of large gauge invariance
was not recognized. This is the only place where explicit dependence
on colour degrees of freedom appear. This coupling introduces a colour
source into the problem for a fixed $A_0$ field and projection onto
the colour neutral states by integrating over the $A_0$ field, in a
gauge invariant manner, as required. Actually, at the one quark loop
level there is an accidental $Z(N_c)$ symmetry in the model which
generates a similar selection rule as in pure gluodynamics, from which
a strong thermal suppression, $ {\cal O} (e^{-N_c M /T } )$ follows.
In this way compliance with ChPT can be achieved since now $ L_i^* -
L_i \simeq e^{-N_c M / T}$ but also puts some doubts on whether chiral
quark models still predict a chiral phase transition at realistic
temperatures. This question has been addressed using specific
potentials for the Polyakov loop either based on one loop perturbation
theory for massive gluons~\cite{Meisinger:2002kg} in the high
temperature approximation or on strong coupling expansions on the
lattice~\cite{Fukushima:2003fw}. In both cases similar mean field
qualitative features are displayed; the low temperature evolution is
extremely flat, but there appears a rapid change in the critical
region, so that $ \langle \bar q q \rangle^* \simeq \langle \bar q q
\rangle $ when $ \langle \Omega \rangle \simeq 0 $ and $ \langle \bar
q q \rangle \simeq 0 $ when $ \langle \Omega \rangle \simeq 1 $.  A
more general discussion and diagramatic interpretation of these issues
as well as the influence of higher quark loop effects and dynamical
Polyakov loop contributions will be presented
elsewhere~\cite{Megias:2004} providing a justification of the one
quark loop approximation at least at low temperatures. There one
obtains that the Polyakov loop effect can be factored out as
follows\footnote{Note that in this formula $\langle \bar q ( n \beta )
q( 0) \rangle$ refers to quarks uncoupled to the Polyakov loop while
in Eq. (\ref{eq:sel}) it refers to quenched QCD.}
\begin{eqnarray}
\langle \bar q q \rangle^* &=& \sum_{n} \frac1{N_c} {\rm
Tr}_c ((-\Omega)^n )\langle \bar q ( n \beta ) q( 0) \rangle \,.
\end{eqnarray} 
This result is consistent with applying the center symmetry selection
rule, Eq.~(\ref{eq:sel}), to the $Z(N_c)$ breaking condensate,
Eq.~(\ref{eq:cond_njl}), of the chiral quark model without Polyakov
loops. If one now takes a suitable average on Polyakov loop
configurations consistent with center symmetry, i.e., including for
each such configuration all its Gribov replicas, Eq. (\ref{eq:Pave})
applies. Schematically, this yields
\begin{eqnarray}
\langle \bar q q \rangle^* &\sim& \sum_{n} \langle \bar q ( n N_c
\beta ) q( 0) \rangle
\sim  \sum_{n} e^{-n N_c M / T} 
\end{eqnarray}
in the confining phase. (In the above sums each term carries a weight
coming from the Polyakov loop average and phase space factors.)  On
the other hand in the unconfining phase, where the center symmetry is
spontaneously broken, the Polyakov loop is nearly unity and one
recovers the standard chiral quark models results, without Polyakov
loop coupling.

\section{Chiral Lagrangians at finite temperature}

It is interesting to construct the coupling of Polyakov loops with
composite pion fields at finite temperature. Using the heat kernel
techniques presented in Ref.~\cite{Megias:2002vr} and already applied
to massless QCD ~\cite{Megias:2003ui}, we can obtain the lowest order
chiral Lagrangian
\begin{eqnarray}
\cL_q^{(2)} &=& \frac{f^*_\pi{}^2}{4}\tr_f\left( \D_\mu U^\dagger\D_\mu
U +(\overline\chi^\dagger U +\overline\chi U^\dagger) \right)
\end{eqnarray} 
where $U$ is the non-linear transforming pseudoscalar Goldstone field,
$ \bar \chi $ the quark mass matrix and $\tr_f $ is the trace in
flavor space. The pion weak decay constant, $f_\pi^*$, at finite
temperature in the presence of the Polyakov loop and in the chiral
limit is given by
\begin{eqnarray}
f^*_\pi{}^2 &=& 4  M^2 \, T \, {\rm Tr}_c \sum_{\hat \omega_n} \int \frac{d^3
k}{(2\pi)^3} \frac1{\left[\hat \omega_n^2+ k^2 + M^2 \right]^2} \nonumber \,.
\end{eqnarray}
The full calculation of the low energy constants at order ${\cal
O}(p^4) $ as a function of temperature and the Polyakov loop is
carried out in Ref.~\cite{Megias:2004}. The main feature is, similarly
to $ \langle \bar q q \rangle^* $ and $ f_\pi^* $, a strong
suppression ${\cal O } ( e^{-N_c M \beta }) $ at low temperatures, but
an enhancement of quark thermal effects close to the
chiral-deconfinement phase transition. 

\section{Conclusions} 

We see that the coupling of the Polyakov loop to chiral quark models
at finite temperature accounts for large gauge invariance and modifies
in a non-trivial way the results for physical observables. On the one
hand, such a coupling allows to satisfy the requirements of chiral
perturbation theory at low temperatures, generating a very strong
suppression at low temperatures of quark loop effects.  Nonetheless,
the onset of deconfinement through a non vanishing value of the
Polyakov loop accounts for a chiral phase transition at somewhat
similar temperatures as in the original studies where the Polyakov loop
was set to one. We expect this feature to hold also in the calculation
of other observables. Although these arguments do not justify by
themselves the application of these chiral quark-Polyakov models to
finite temperature calculations, they do show that they do not
contradict basic expectations of QCD at finite temperature.

{\sl This work is supported in part by funds provided by the Spanish DGI
with grant no. BMF2002-03218, Junta de Andaluc\'{\i}a grant no. FM-225
and EURIDICE grant number HPRN-CT-2003-00311.} 

\end{document}